# A Neural Network-Based Submesoscale Vertical Heat Flux Parameterization and Its Implementation in Regional Ocean Modeling System (ROMS)

**Shuyi Zhou[1], Jihai Dong[2,3], Fanghua Xu[1], Zhiyou Jing[4], and Changming Dong[2,3]**

[1] Department of Earth System Science, Ministry of Education Key Laboratory for Earth System Modeling, Institute for Global Change Studies, Tsinghua University, Beijing, 100084, China.

[2] School of Marine Sciences, Nanjing University of Information Science and Technology, Nanjing, Jiangsu, China.

[3] Southern Marine Science and Engineering Guangdong Laboratory (Zhuhai), Zhuhai, Guangdong, China.

[4] State Key Laboratory of Tropical Oceanography, South China Sea Institute of Oceanology,

Chinese Academy of Sciences, Guangzhou, China

Corresponding author: Jihai Dong (jihai_dong@nuist.edu.cn)

**Key Points:**

- We propose a neural network-based submesoscale vertical heat flux parameterization which is trained on ultra-high resolution model data.
- The new scheme can accurately capture the submesoscale vertical heat flux and characterize how it varies with depth.
- The new scheme improves the capability of the regional model in simulating mesoscale dynamics and mixed layer depth.

**Abstract**

Submesoscale processes, with spatio-temporal scales of $O(0.01\text{-}10)$ km and hours to 1 day which are hardly resolved by current ocean models, are important sub-grid processes in ocean models. Due to the strong vertical currents, submesoscale processes can lead to submesoscale vertical heat flux (SVHF) in the upper ocean which plays a crucial role in the heat exchange between the atmosphere and the ocean interior, and further modulates the global heat redistribution. At present, simulating a submesoscale-resolving ocean model is still expensive and time-consuming. Parameterizing SVHF becomes a feasible alternative by introducing it into coarse-resolution models. Traditionally, researchers tend to parameterize SVHF by a mathematically fitted relationship based on one or two key background state variables, which fail to represent the relationship between SVHF and the background state variables comprehensively. In this study, we propose a data-driven SVHF parameterization scheme based on a deep neural network and implement it into the Regional Ocean Modeling System (ROMS). In offline tests, our scheme can accurately calculate SVHF using mesoscale-averaged variables and characterize how it varies with depth. In online tests, we simulate an idealized model of an anticyclonic mesoscale eddy and a realistic model of the Gulf Stream, respectively. Compared to the coarse-resolution cases without the SVHF effect, the coarse-resolution cases with the SVHF scheme tend to reproduce results closer to the high-resolution case and the observational state in terms of the temperature structure and mixed layer depth, indicating a good performance of the neural network-based SVHF scheme. Our results show the potential of applying the neural network in parameterizing sub-grid processes in ocean models.

**Plain Language Summary**

The vertical heat transport in the ocean is crucial for studies of climate change and global warming. In the last decades, oceanographers have dedicated themselves to quantifying the heat budget in the ocean, especially the heat transport by different dynamic processes such as thermohaline circulation, mesoscale eddies, and micro-scale mixing. Recently, submesoscale processes have been found to make a significant contribution to the vertical heat transport due to its ageostrophy. Due to the small spatial scale of submesoscale processes, the submesoscale-induced vertical heat flux is hardly reproduced by ocean models, which limits the capability of the ocean and climate models to simulate the heat budget. Here, we train a data-driven neural network to parametrize the submesoscale vertical heat flux by using ultra-high resolution model data and implement it in a coarse-resolution model. The new scheme is demonstrated to have a good performance in improving the model simulating capability.

**1 Introduction**

The vertical transport in the ocean is of significant for the global heat budget and the carbon cycle. The geostrophy limits vertical motions in the ocean and geostrophic currents can only induce negligible vertical transports. Because of the small spatio-temporal scales of $O(0.01\text{-}10)$ km and hours to 1 day (McWilliams, 2016) which breaks the geostrophy, studies have claimed that submesoscale processes in the upper ocean can lead to strong vertical currents up to $O(100)$ m day$^{-1}$, and is a crucial routine of mass, energy, and biogeochemical exchanges within the ocean in vertical and further affect the exchanges between the ocean and the atmosphere (Lévy et al., 2012; McWilliams, 2016; Siegelman et al., 2020a; Yu et al., 2019; Zhengguang Zhang & Qiu, 2018). However, the small spatial scale makes it difficult for current ocean models, especially global climate models, to reach the resolution required for resolving submesoscale processes and reproducing the induced vertical transports as time-consuming and expensive costs (J. Dong et al., 2020). Therefore, it is a better choice to parameterize the vertical transport by submesoscale processes in current coarse-resolution models.

Frontogenesis and mixed layer baroclinic instability are two major mechanisms generating submesoscale processes in the ocean surface mixed layer (MLD) (McWilliams, 2016; Taylor & Thompson, 2023). Frontogenesis and instability theories both predict that the secondary circulations of submesoscales tend to result in upward heat flux (Boccaletti et al., 2007; Fox-Kemper et al., 2008; Hoskins, 1982). The vertical heat flux induced by submesoscale processes is 5 times as many as that of mesoscale processes (Su et al., 2018) and of comparable magnitude to air–sea fluxes (Siegelman et al., 2020b). Submesoscale processes are one of the crucial counter-gradient vertical heat flux mechanisms (Deardorff, 1966; Priestley & Swinbank, 1947) which can inhibit the growth of La Niña and El Niño events (S. Wang et al., 2022). This upward heat flux is also accompanied by the restrafication of the MLD (Bachman & Taylor, 2014). It has been demonstrated that submesoscale-resolving models tend to simulate the MLD thickness closer to the observed one than the coarse resolution models (Ding et al., 2022; Treguier et al., 2023).

Several schemes have been proposed to parameterize the submesoscale vertical heat flux (SVHF) in ocean models. By introducing an overturning streamfunction, Fox-Kemper et al. (2008) proposed a submesoscale scheme that can parameterize the SVHF and has been widely implemented in ocean models (Calvert et al., 2020; Fox-Kemper et al., 2011). In the parameterized express, the SVHF is crucially determined by the horizontal buoyancy gradient resolved by the model and also the mixed layer depth. As only the SVHF by mixed layer instability (MLI) is scaled in the scheme, a new scheme is further proposed by Zhang et al. (J. Zhang et al., 2023) which considers SVHF from both MLI and frontogenesis. Another important difference is that the scheme by Zhang et al. (2023) allows the SVHF to penetrate the mixed layer base which is closer to observations (Zhang et al., 2021). Besides, Xuan et al. (2019) find the total vertical advection in the high-resolution model is an order of magnitude higher than that of the low-resolution one. It is therefore proposed to parameterize SVHF by increasing the vertical thermal diffusion coefficient by $2cm^2/s$ in low-resolution models, which may be only regionally workable, and not be applicable universally. In summary, current SVHF schemes are almost proposed based on traditional methods, and one always should have to fit a mathematical relationship empirically by choosing a few key factors related to the background fields. However, submesoscale processes can be generated from different mechanisms, and these expressions may not represent the relationship between the SVHF and the background fields comprehensively.

The rapid development of deep learning provides a new path for the study of atmospheric and oceanic subgrid dynamic process parameterization (C. Dong et al., 2022). On the one hand, neural network-based modeling can skip the complicated physical process. On the other hand, it has a strong ability to learn nonlinear relationships (Liu, 2012) and is very suitable for describing the complex relationship between mean state variables and the SVHF. Up to now, neural networks have been used to carry out many studies in marine turbulence parameterization, such as subgrid eddy momentum forcing (Bolton & Zanna, 2019; Guillaumin & Zanna, 2021; Perezhogin et al., 2023; C. Zhang et al., 2023), ocean mesoscale eddies (Perezhogin & Glazunov, 2023; Zanna & Bolton, 2020), subgrid-scale fluxes (Srinivasan et al., 2023), turbulent diffusion coefficient (Liang et al., 2022; Mashayek et al., 2022). Zhu et al. (2022) propose a vertical mixing parameterization based on a neural network with physical constraints, which is trained by turbulence observational data in the tropical Pacific Ocean. Compared to traditional parameterization in the Ocean general circulation model (OGCM), the data-driven turbulent mixing parameterization can better simulate the vertical heat flux in the upper ocean, thereby improving the temperature simulation results in the tropical Pacific Ocean. The vertical mixing parameterization can also be improved by enhancing its eddy diffusivity model (Sane et al., 2023), which reduces the mixing layer and thermocline deviation in the OGCM. Compared with traditional parameterization, parameterization based on the neural network contains multiple mechanisms and is no longer the description of a single mechanism. In addition to using observational data, data from high-resolution models or large eddy simulations is also a good option. Bodner et al. (2023) first utilize the Convolutional Neural Network trained by LLC4320 to parameterize the vertical buoyancy flux of the submesoscale process in the ocean mixed layer, which has better performance than the physically-based scheme. However, this scheme cannot describe the change of submesoscale vertical buoyancy flux with depth, which cannot directly be implemented in an ocean model.

In this study, we propose a parameterization of the SVHF using neural networks and evaluate its performance in simulations of an idealized anticyclonic mesoscale eddy and the realistic Gulf Stream region. We aim to accurately capture the relationship between the SVHF and other variables using neural networks, and incorporate it into the low-resolution ocean model. The rest of the paper is organized as follows: Section 2 introduces the training data, pre-processing and the neural network structure, and how the SVHF is parameterized, Section 3 evaluates the performance of the new parameterization in the offline test, Sections 4 and 5 show the effects of the new parameterization in simulations of an idealized anticyclonic mesoscale eddy and Gulf Stream, respectively. Finally, discussion and conclusions are given in Section 6.

## 2 Data and Methods

### 2.1 Training Dataset and Pre-processing

The training dataset in this study utilizes the output from a model simulated by the Regional Ocean Modeling System (ROMS; Shchepetkin & McWilliams, 2005) with an ultra-high spatial resolution of 500m. Figure 1a illustrates the model configuration which is a three-layer offline nesting: the parent domain ROMS0, the intermediate nested layer ROMS1, and the most inner layer ROMS2. The horizontal resolutions of ROMS0, ROMS1, and ROMS2 are 7.5km, 1.5km, and 0.5 km, respectively. Vertically, the model has 60 layers with more layers concentrating within the MLD. The parent ROMS0 is run for 22 years and the last 1-year data are used for the intermediate ROMS1. As submesoscales are more active in winter, the ROMS2 is run from January to February, which can well capture the submesoscale processes and energy transfer in this region (Cao et al., 2021; Luo et al., 2020). A more detailed model configuration can be found in Cao et al. (2021) and Dong et al. (2022). The spatial distribution of the normalized relative vertical vorticity ($\zeta/f$; an indicator of submesoscale processes which is active when $\frac{\zeta}{f} \sim 1$) depicted in Figure 1b reveals pronounced submesoscale processes in the upper layer. As the SVHF can extend below the mixed layer base (J. Zhang et al., 2023), we select the data from the sea surface to the depth of 1.2 times MLD as the training dataset. Here, we employ a temperature difference of 0.2 °C to determine the MLD, as it is simple and stable (de Boyer

Montégut et al., 2004). More specifically, the data is subsampled at a horizontal interval of 1 km to reduce the training cost. In this study, the SVHF is defined as

$$SVHF = \overline{w'T'} \tag{1}$$

where $w$ and $T$ are the vertical velocity and temperature. The prime denotes the submesoscale component which is obtained by applying a spatial highpass filter with a cutoff scale of 20 km (this scale is close to the local deformation radius of the MLI; cf. Dong et al., 2020), and the overbar denotes spatially averaging over a 20-km box. Observations and simulations have shown that the vertical structure of the SVHF is tightly related to the MLD, so a normalized depth by the MLD is used to avoid the potential effect of the absolute MLD magnitude,

$$z_n = \frac{z}{h_{MLD}} \tag{2}$$

where $h_{MLD}$ is the mixed layer depth at a certain location, and $z$ is the depth.

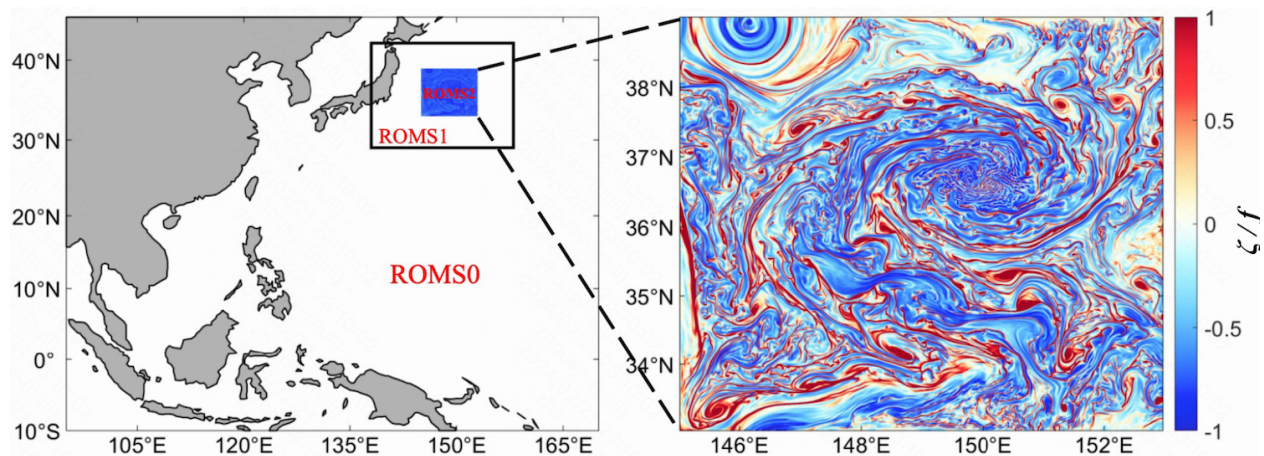

**Figure 1**. (a) Location of the model domain. The whole domain is ROMS0, the black box is ROMS1 and the inner domain is ROMS2 with color contours showing a snapshot of the sea surface normalized relative vorticity ($\zeta/f$) on 1 February and a zooming view showing in (b).

## 2.2 Neural Network-Based SVHF Parameterization

In this study, we use the deep neural network (DNN) to construct the SVHF parameterization, which has been widely used in the development of ocean subgrid process parameterization and successfully applied in ocean models. Herein, the DNN has a five-layer structure, including an input layer, an output layer, and three hidden layers. The number of neurons in the three hidden layers is 64, 16, and 8, respectively. The three hidden layers have been able to fit most of the formulas (Hornik et al., 1989) and the relatively simple model also can reduce the calculation overhead in the online test (Sane et al., 2023). We employ the LeakyRule as the activation function to characterize the nonlinear relationship between the mean state variables and the SVHF. During the model training, we use the mean square error as the loss function, and adopt the early stopping (patience = 30) to prevent overfitting. The dataset is divided into the training data and the test dataset by a ratio of 7:3. What's more, all variables are standardized uniformly,

$$\hat{x} = \frac{x - \mu}{\sigma}, \tag{3}$$

where $x$ is the variable to be standardized, $\hat{x}$ is the standardized variable, $\mu$ denotes the mean value of $x$, and $\sigma$ denotes the standard deviation of $x$. Figure 2 shows the overall structure of DNN. The DNN adopts the quantities related to the background state variables including temperature ($\bar{T}, \frac{\overline{\partial T}}{\partial x}, \frac{\overline{\partial T}}{\partial y}$), the horizontal velocities ($\bar{u}, \frac{\overline{\partial u}}{\partial x}, \frac{\overline{\partial u}}{\partial y}, \bar{v}, \frac{\overline{\partial v}}{\partial x}, \frac{\overline{\partial v}}{\partial y}$), and $z_n$ as input, and the SVHF ($\overline{w'T'}$) as output. The relationship is shown in the following formula,

$$\overline{w'T'} = F\left(\bar{T}, \frac{\overline{\partial T}}{\partial x}, \frac{\overline{\partial T}}{\partial y}, \bar{u}, \frac{\overline{\partial u}}{\partial x}, \frac{\overline{\partial u}}{\partial y}, \bar{v}, \frac{\overline{\partial v}}{\partial x}, \frac{\overline{\partial v}}{\partial y}, z_n\right) \quad (4)$$

where $\frac{\partial}{\partial x}$ and $\frac{\partial}{\partial y}$ denote the meridional and zonal gradients of variables, respectively. Here, different from the traditional parameterization schemes, the effect of the latitude (i.e., the Coriolis parameter) is not considered in the expression. The data used for training here cover a small range meridionally, which may not be enough to capture the sensitivity of the SVHF to the Coriolis parameter. Meanwhile, the vertical shear of the horizontal currents is also not taken as input, since the vertical shear is dynamically equal to the horizontal buoyancy gradient under geostrophic balance which should be mostly reflected in the horizontal temperature gradient. The essence of the neural network calculation lies in the amalgamation of matrix multiplication and activation functions (Equation 5). Therefore, the parameters in each layer within the DNN can be imported into the ocean model and participate in the calculation process at every step,

$$y_i = LeakyReLU(w_i x_i + b_i). \quad (5)$$

where $y_i$ is the output, $x_i$ is the input, $w_i$ is the weight, and $b_i$ is the bias. The advantage of the DNN is that all variables that potentially modulate the SVHF are considered. Despite that some quantities make limited contributions (see the section below), the comprehensive consideration potentially improves the capability of the scheme to reproduce the SVHF.

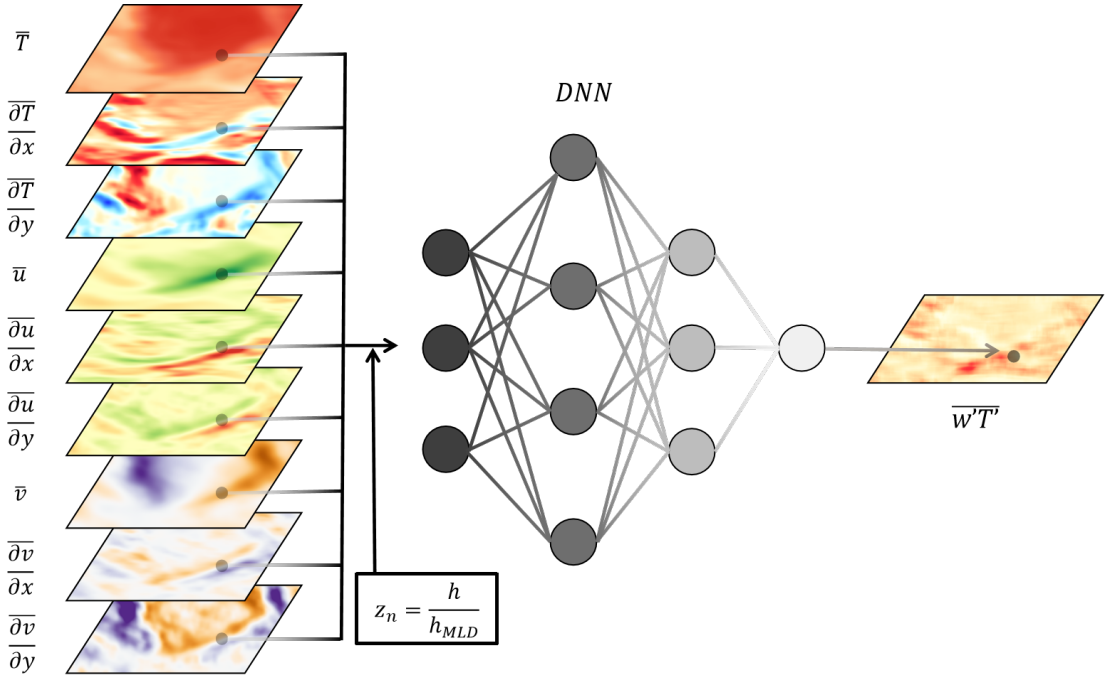

**Figure 2**. Overall architecture of the DNN. The DNN adopts nine mean state variables ($\bar{T}, \frac{\overline{\partial T}}{\partial x}, \frac{\overline{\partial T}}{\partial y}, \bar{u}, \frac{\overline{\partial u}}{\partial x}, \frac{\overline{\partial u}}{\partial y}, \bar{v}, \frac{\overline{\partial v}}{\partial x}, \frac{\overline{\partial v}}{\partial y}, z_n$) as input, and the SVHF ($\overline{w'T'}$) as output.

## 3 Offline Test

After the new scheme based on the DNN model is established, the scheme performance is firstly assessed by comparing it to the directly calculated SVHF from ROMS2 (the ground truth hereinafter; Figure 3). During the comparison, the data for training are excluded. As shown in Figure 3a, the ground truth of the depth-averaged SVHF within a normalized depth of 1.2 (i.e., $1.2h_{MLD}$) is dominated by positive fluxes over the whole region, which is consistent with prior studies (Su et al., 2018). In vertical, the meridionally-averaged SVHF is mainly concentrated within the MLD but can penetrate the ocean interior below the mixed layer base (Figure 3d). Overall, a quantitative comparison shows that the DNN model generally capture the SVHF patterns horizontally and vertically (Figure 3d, e *vs.* Figure 3a, b). To evaluate the DNN model performance more quantitively, the bivariate histogram of the depth-averaged SVHF is shown (Figure 3c) which is mainly concentrated around the diagonal, indicating that the DNN can statistically reproduce the SVHF accurately. The average bias in the test dataset is $-7.7 \times 10^{-7}\ m \cdot s^{-1} \cdot °C$, which is two orders of magnitude smaller compared to the SVHF magnitude. The domain-averaged SVHF shows notable vertical variations (red line in Figure 3f). As depth increases, the SVHF increases sharply first and then decreases, peaking at about 40m. It is noteworthy that the SVHF is still considerable below the mixed layer base, which is consistent with observations (Zhiwei Zhang et al., 2021). By comparison, despite that the DNN-based scheme tends to slightly underestimate the SVHF peak but overestimate it below, the scheme well reproduces the variation of the SVHF, especially the increasing trend of the SVHF above the peak (blue line vs. red line) and the peak depth. Overall, the consistency demonstrates the capability of the DNN-based scheme to reproduce the SVHF.

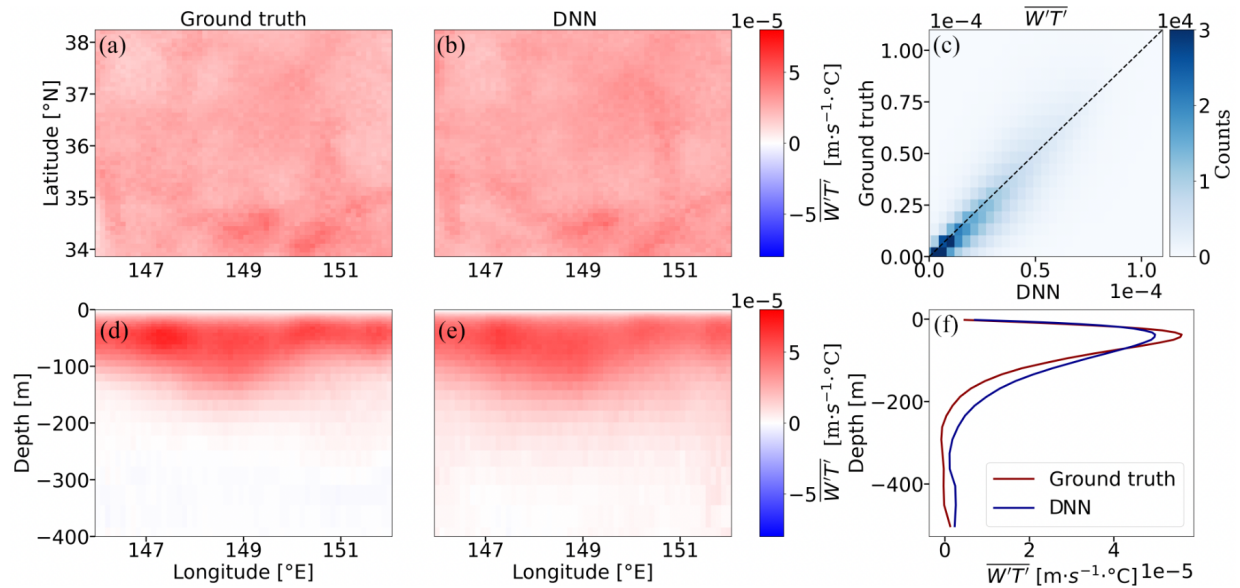

**Figure 3**. Comparison of the SVHF between the DNN-based scheme and the ground truth. The depth-averaged (a) ground truth and (b) the DNN results, and (c) their bivariate histograms. The meridionally-averaged (d) ground truth and (e) DNN results, and (f) their domain-averaged vertical profiles. The colors in (c) represent the number of data points in each bin.

We conduct permutation experiments (Zhou et al., 2022) to further investigate the importance of different variables in this scheme. The experiment evaluates the variable importance by comparing the change in the accuracy of the output after a certain variable is out-of-order. Here, we use the relative importance index (RiI) to measure the variable importance,

$$RiI = \frac{RMSE_p - RMSE}{RMSE} \quad (5)$$

where $RMSE$ is the original root mean square error between output and ground truth, and $RMSE_p$ is the $RMSE$ after a certain variable is out-of-order.

Figure 4 illustrates the RiI values of different variables. The larger the RiI, the more important the variable is. Note that, $\frac{\overline{\partial u}}{\partial y}$, $\frac{\overline{\partial v}}{\partial x}$, and $z_n$ have the largest RiI. As the SVHF is sensitive to the vertical depth (Figure 3f), it is reasonable that $z_n$ is crucial for the parameterization to accurately calculate *SVHF* at different depths. Furthermore, $\frac{\overline{\partial u}}{\partial y}$ and $\frac{\overline{\partial v}}{\partial x}$ can be subtracted to form vorticity ($\zeta = \frac{\overline{\partial v}}{\partial x} - \frac{\overline{\partial u}}{\partial y}$), which is a reflection of the intensity of submesoscale processes and potentially modulates the magnitude of the *SVHF*. The order of variable importance in different regions may change with local temperature and current characteristics. The Kuroshio Extension has strong zonal flow, where the changes of $\frac{\overline{\partial u}}{\partial x}$ and $\frac{\overline{\partial v}}{\partial y}$ are small. As a result, the model is not sensitive to their changes.

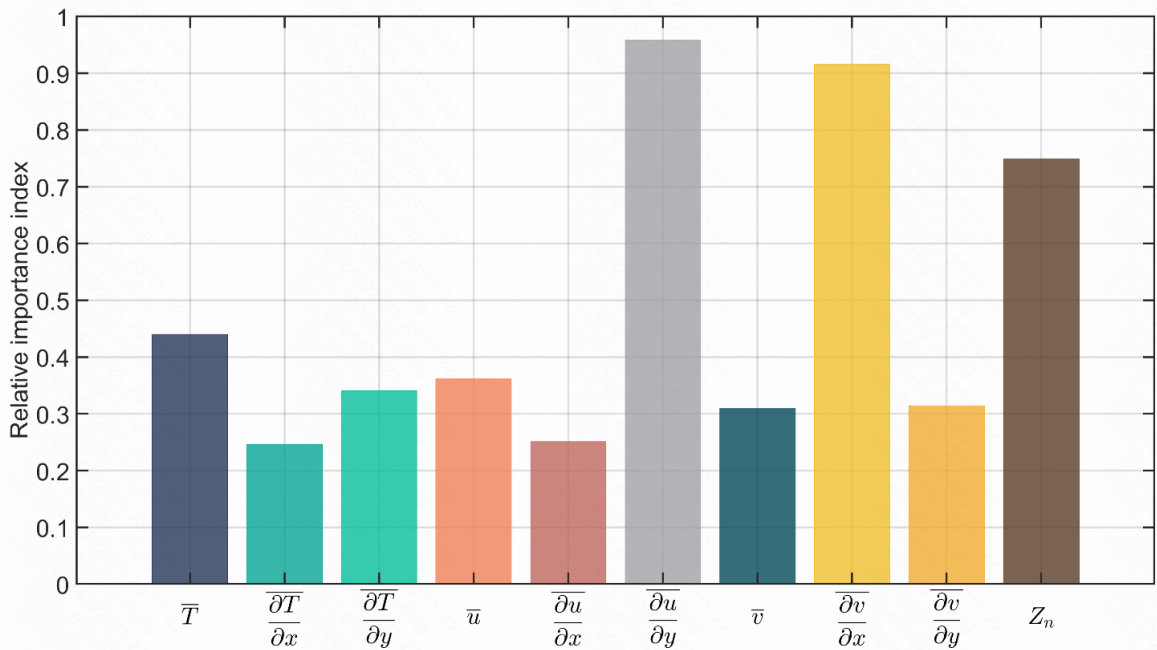

**Figure 4**. The relative importance index (RiI) of each predictor in the DNN model. The lager RiI, the more important the variable is.

### 4 Online Test: Idealized Mesoscale Eddy Simulation

Our ultimate goal is to implement it in ocean models, so it is not enough for the DNN-based scheme to only carry out offline tests. The evaluation metrics in the offline test do not necessarily indicate a good performance in online simulations. In this section, we introduce the new scheme into a low-resolution ROMS model to simulate an idealized anticyclonic mesoscale eddy. The eddy evolution in the DNN-scheme-implemented case is expected to be closer to the high-resolution model that resolves submesoscales, compared to the low-resolution one without the SVHF effect. The model configuration is described in Subsection 4.1 and Subsection 4.2 analyses the model result.

## 4.1 Model setup

We simulate an idealized anticyclonic mesoscale eddy using ROMS. Referring to the configuration of Wang et al., (2022), we use the climatological temperature in the Kuroshio Extension as the background state and construct an idealized anticyclonic mesoscale eddy to generate submesoscale processes for the SVHF study, as shown in Figure 5. To our knowledge, anticyclonic mesoscale eddies are more conducive to generating submesoscale eddies than cyclonic ones (Brannigan et al., 2017). The study domain is 600 km × 600 km. The model has 60 layers in vertical with a flat bottom of 1000m depth. The Coriolis parameter is set to be constant ($f = 10^{-4}$; a value around the KE region). To highlight the new scheme performance, three simulations are conducted: a 1-km one (the truth case which can resolve submesoscales and directly simulates the SVHF; CTRL-1km hereinafter), a 5-km one with the DNN model activated (the SVHF is represented by the DNN model; DNN-5km hereinafter), and a 5-km one without the SVHF effect (CTRL-5km hereinafter). All simulations are run for 180 days to allow the evolution of the eddy and the development of submesoscales. For simulating efficiency and stability, the DNN-based scheme is triggered when the depth is less than 400 m and the strain rate is greater than 0.25.

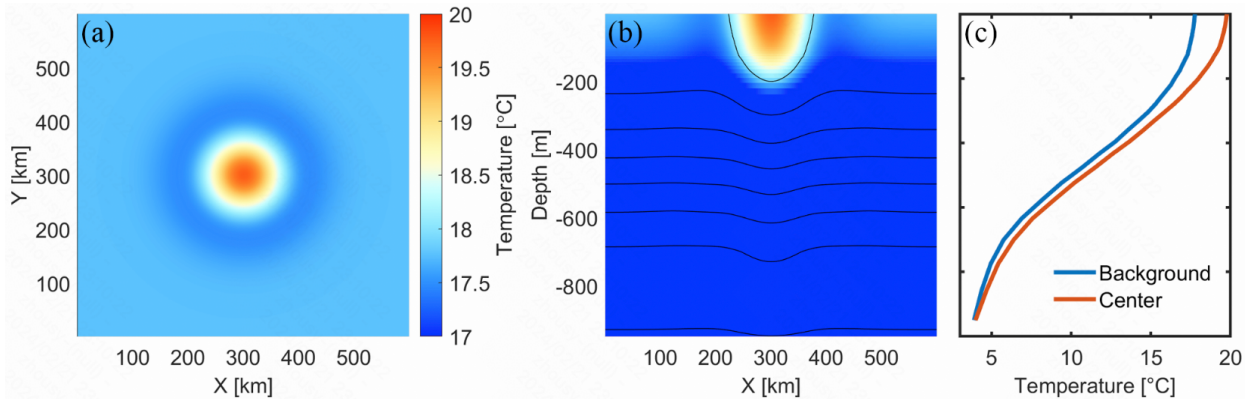

**Figure 5**. The initial temperature field of the idealized mesoscale eddy: (a) the plane view from the sea surface; (b) the lateral view across the eddy center; (c) the vertical temperature profiles at the eddy center (orange line) and out of the eddy (blue line).

## 4.2 Effect of the DNN Scheme on Eddy Structure and Energy

Before 140 days, there is little difference in the pattern of sea surface temperature and zonally-averaged temperature profile among these three cases (Figure 6). After 140 days, submesoscale processes become active, and differences in the morphology of the mesoscale eddy between the cases become increasingly obvious. On Day 160, the eddy structure in CTRL-1km begins to be destroyed, and submesoscale eddies at the edge of the eddy begin to form. Although small eddies could be also seen around the eddy in the low-resolution simulations, CTRL-5km and DNN-5km, the intensity and activity are much weaker. Nevertheless, compared to CTRL-5km in which the eddy is finally split into two smaller eddies, the SVHF effect makes the eddy structure in DNN-5km resemble that in CTRL-1km and both have only one dominant eddy. From the lateral sections across the eddy, because the eddy in CTRL-5km is split into two eddies, the isothermal is depressed twice but the displacements are small. By contrast, the eddies in both CTRL-1km and DNN-5km depress the isothermal much deeper, suggesting the positive effect of the DNN-based scheme.

The differences in the isothermals among the cases suggest the modulation by the SVHF on the energy budget. Figure 7 shows the lateral sections of the meridionally-averaged available potential energy (APE). On Day 140, the APE patterns in different cases are similar. The difference between CTRL-5km and the other two cases begins to emerge after Day 160, with two high-value regions in the distribution of APE corresponding to the two eddies. The difference is even more pronounced on Day 180. Generally, the simulation of APE in the DNN-5km is improved due to the SVHF effect and is much closer to the true case CTRL-1km, in contrast to CTRL-5km. Apart from APE, the simulation of kinetic energy (KE) is also improved (Figure 8). KE in CTRL-5km is significantly different from the other two cases on Day 180, the KE patterns in DNN-5km are similar to CTRL-1km on different days. From the perspective of energy, it is further explained that the performance of the coarse-resolution model can be improved by including the SVHF effect via implementing the DNN scheme.

## 5 Online Test: Gulf Stream Simulation

The upward SVHF makes a significant contribution to the restratification of the mixed layer. Recent studies report that current ocean models with low (i.e., submesoscale-unresolved) horizontal resolutions tend to simulate a deeper MLD than the observed one in many regions, such as the Kuroshio Extension and the Gulf Stream (Ding et al., 2022; Treguier et al., 2023). This may be due to a lack of submesoscale processes. The DNN is trained based on the simulation data in the Kuroshio Extension. Therefore, in this section, we simulate the Gulf Stream using ROMS to investigate the effectiveness of the DNN scheme. Section 5.1 is the model configuration, and Section 5.2 shows the model results.

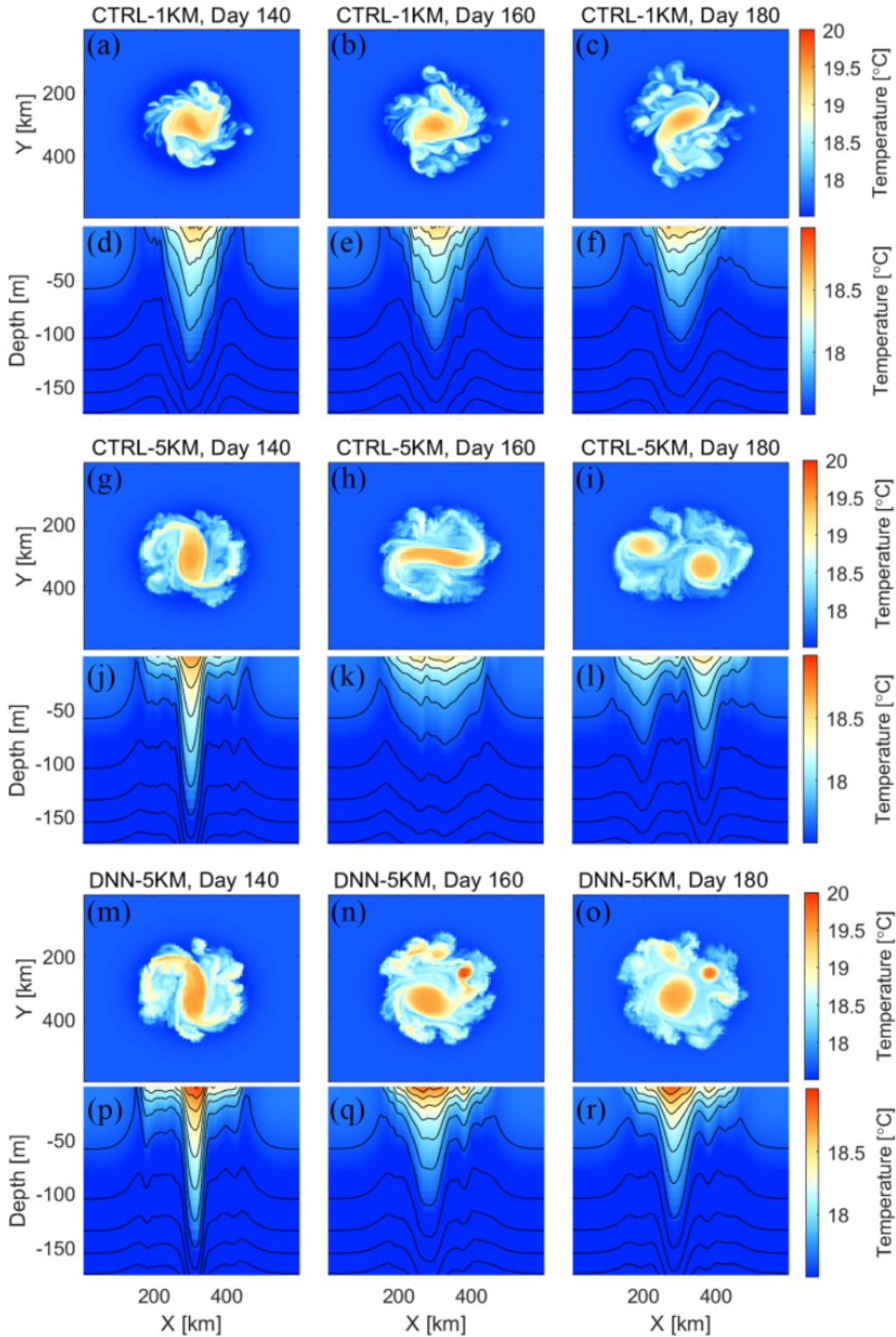

**Figure 6**. Snapshots of the sea surface temperature (a-c, g-i, m-o) and the meridionally-averaged temperature (d-f, j-l, p-r) sections on Day 140 (left column), 160 (middle column), and 180 (right column). The upper two rows are from CTRL-1km, the middle two are from CTRL-5km, and the lower two are from DNN-5km.

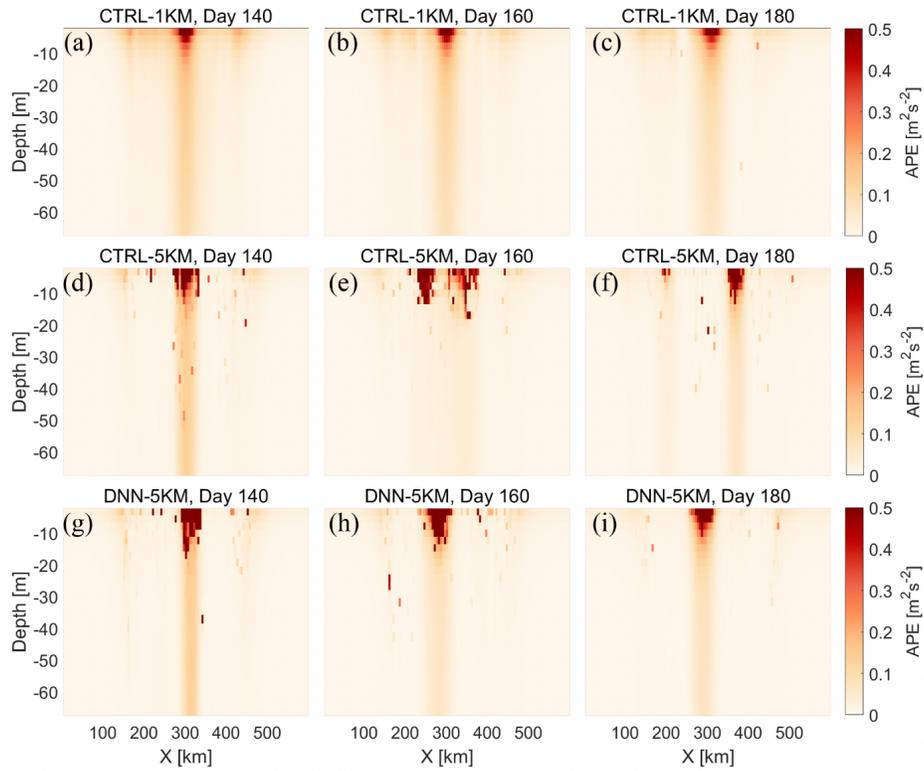

**Figure 7**. Snapshots of the meridionally-averaged APE sections from (a, b, c) CTRL-5km, (d, e, f) CTRL-1km and (g, h, i) DNN-5km on Day 140 (left column), 160 (middle column), and 180 (right column).

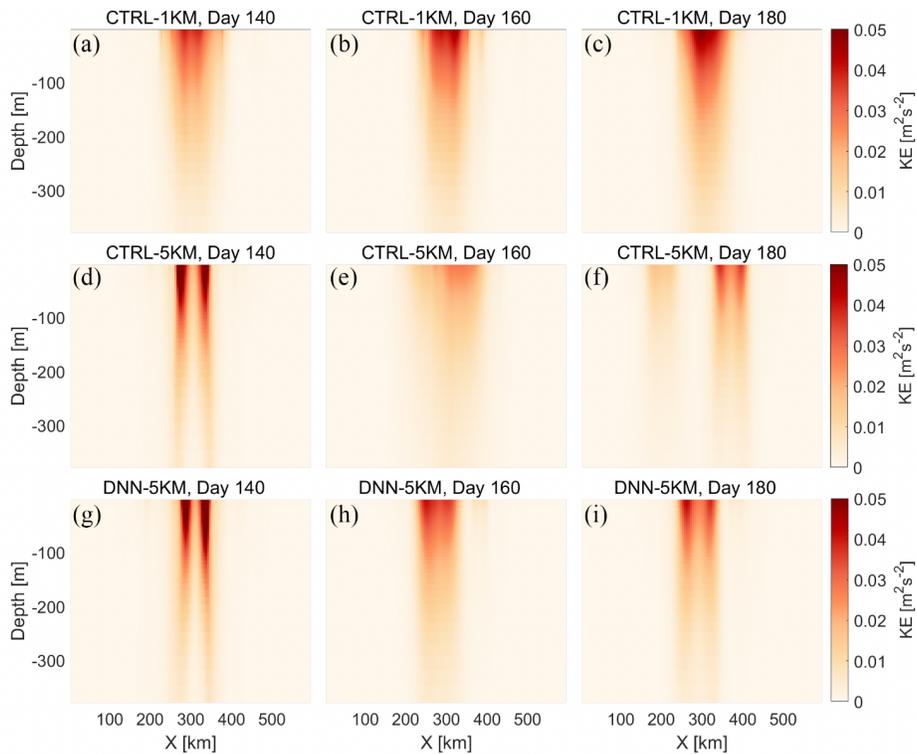

**Figure 8**. The same as Figure 7 but for the kinetic energy.

## 5.1 Model setup

The model is still established by using ROMS with a domain of 27-47°N, 108-145°E. The horizontal resolution of the model is 1/8°, which is mesoscale-resolving. Vertically, the model has 60 levels in the terrain-following S-coordinates. We use the temperature, salinity, and velocity from the HYbrid Coordinate Ocean Model (HYCOM) as the initial and boundary information, and wind speed, humidity, pressure, and precipitation from Climate Forecast System Reanalysis (CFSR) as the forcing. Two cases without (named CTRL) and with (named DNN) the DNN scheme are simulated, respectively. For the sake of numerical stability, the DNN scheme is triggered when the water depth is deeper than 500m and the MLD is less than half the water depth. The simulations are run from January to February 2018 and the first month is used as a spin-up. We just use the February-averaged result for comparison.

## 5.2 Effect of the DNN scheme on MLD

Comparing the CTRL run with the DNN one, the pattern of sea surface temperature does not change significantly. Nevertheless, the overall sea surface temperature increases slightly, which is consistent with Fox-Kemper et al. (2011). The temperature around the Gulf Stream mainstream varies strongly, with alternating warm-core and cool-core eddies, where submesoscale processes should be active. As a result, we select the mainstream region (37.5-42.5°N, 112-142°E) to further analyze the effect of the DNN scheme. According to the zonally-averaged section of temperature shown in Figure 9e-f, positive anomalies up to 0.1°C are generally observed in the upper 200 m due to the DNN scheme. The alternating temperature anomaly stripes below 200m may be related to a modulation of the location and strength of mesoscale eddies. Introducing the DNN scheme makes up for the SVHF that is not resolved by the coarse-resolution model and enhances the vertical heat exchange in the upper ocean, which is bound to change MLD. As shown in Figure 9i, due to the restratification of the SVHF, the MLD of the DNN run is mainly shallower than that of the CTRL run.

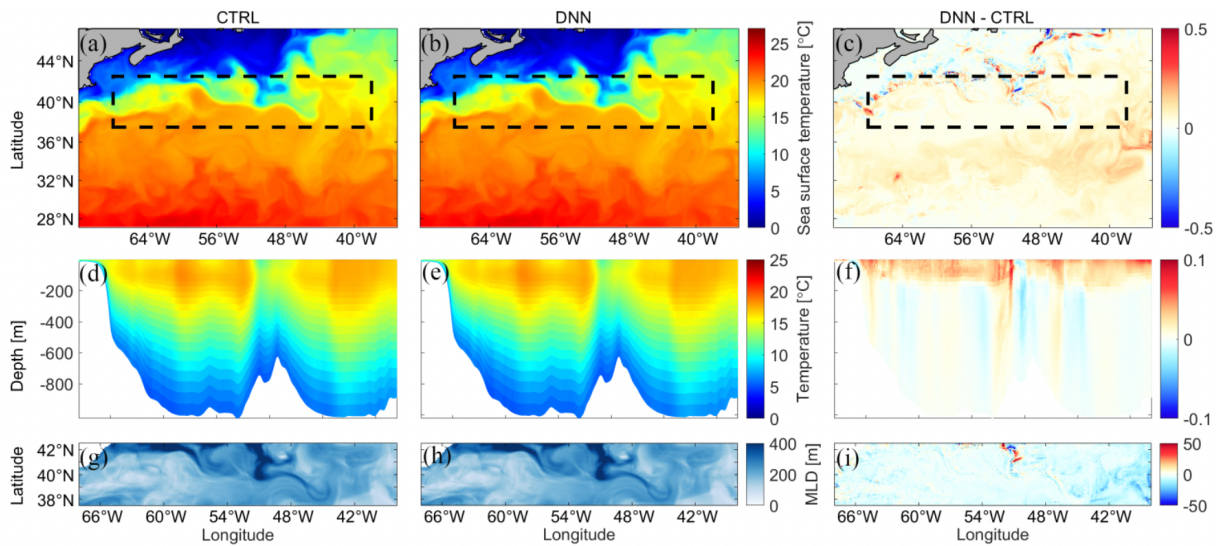

**Figure 9**. The monthly-averaged (a, b) sea surface temperature, (d, e) meridionally-averaged temperature sections and (g, h) MLD in the black box in February 2018 from the CTRL and DNN runs, and (c, f, i) their corresponding differences. To the right column, the bias map is obtained by subtracting CTRL from DNN.

A comparison to the Argo observations shows that the CTRL run tends to simulate a deeper MLD (Figure 10a), which is consistent with Treguier et al. (2023). Influenced by the deep convection in the north of this region, the MLD deepens with increasing latitude. However, the MLD from the DNN run becomes shallower compared to the CTRL run. The MLD can be shallowed by about 4.1m, with a relative reduction of the MLD bias up to 17%. In Figure 10b, the reduction rate of the bias at high latitudes is smaller than that at low latitudes which may be related to the deep water convection in this region (Fox-Kemper et al., 2011). Meanwhile, the bias of DNN is lower than that of CTRL. Overall, it can be observed that the new parameterization can effectively improve the MLD accuracy of the coarse-resolution model.

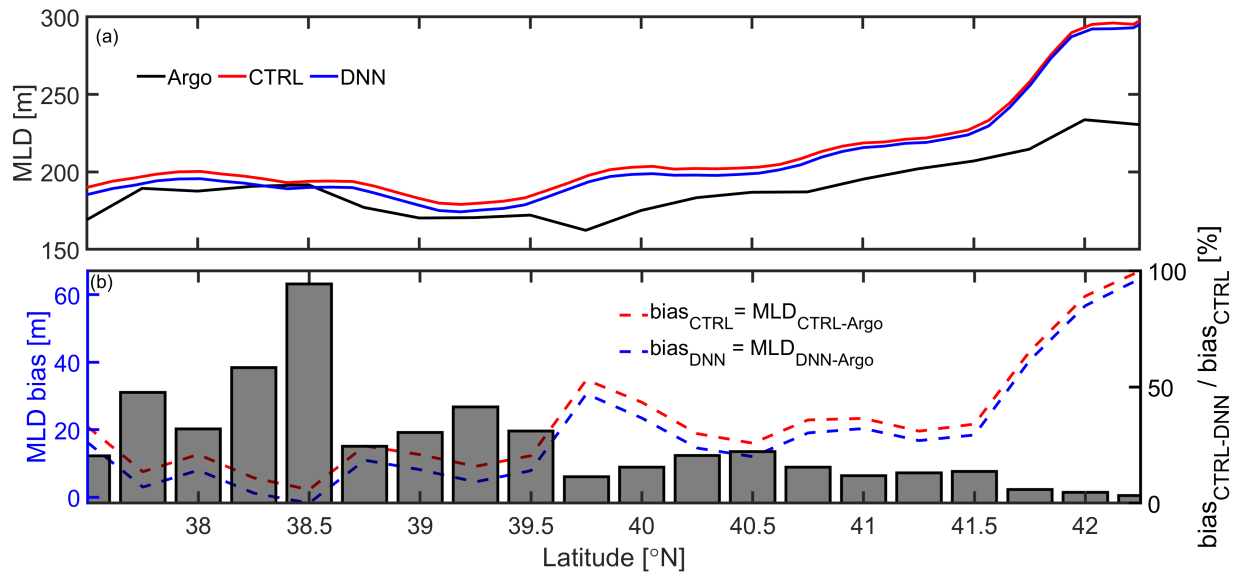

**Figure 10**. Comparison of February-averaged MLD within the black box of Figure 9. (a) The February-averaged MLD of Argo (black line), CTRL (red line), and DNN (blue line) (b) A broken blue line represents the bias of MLD between Argo and DNN, and a broken red line represents the bias of MLD between Argo and CTRL. The gray histogram is the reduction rate of the MLD bias.

## 6 Conclusions and Discussion

In this study, we propose a new SVHF parameterization based on deep neural networks (DNN) and implement it in ROMS. The training dataset is from a submesoscale-resolved ROMS simulation with a horizontal resolution of 500m. In the offline test, the new parameterization can accurately capture the SVHF. Drawing on the idea of shape function (Fox-Kemper et al., 2008), we innovatively introduce a normalized depth by the MLD ($z_n$) as a predictor, which plays a crucial role in accurately characterizing the change of SVHF with depth. In the permutation experiments, we calculate the relative importance of the predictors in the model. Among them, $\frac{\partial \bar{u}}{\partial y}, \frac{\partial \bar{v}}{\partial x}$, and $z_n$ have higher *RiI*. We find that the model learns the vorticity composed of $\frac{\partial \bar{u}}{\partial y}$ and $\frac{\partial \bar{v}}{\partial x}$ which plays a key role in the calculation of the SVHF. In the online test, we successfully implemented the DNN-based scheme to ROMS and simulated two scenarios: one is an idealized anticyclonic mesoscale eddy and the other is a simulation of the Gulf Stream. On the one hand, in the simulation of the idealized eddy, the coarse-resolution eddy with the DNN scheme is closer to the one in the high-resolution model compared to the case without the SVHF effect. On the other hand, adding the DNN scheme reduces the simulation bias of MLD in the Gulf Stream. All these prove that the new scheme is feasible and effective.

According to the submesoscale energy budget equation, the submesoscale vertical buoyancy flux is an important energy conversion routine for the background (i.e., large- and meso-scale potential energy) converting into submesoscale kinetic energy (Cao et al., 2021). The parameterization of the SVHF in the coarse-resolution model partially represents the submesoscale vertical buoyancy flux which tends to characterize a more complete energy budget of the background potential energy. As a result, the simulating capability of the coarse-resolution model is improved by the new scheme.

Although the work here demonstrates that the current neural network-based model can accurately predict SVHF and run stably in the model, it still needs further evaluation and improvement in the future. First, the current training dataset is only from the Kuroshio Extension region from January to March 2018, and the SVHF characteristics of this region cannot represent the whole ocean. However, simulating long-term global submesoscale resolved models is unrealistic and expensive. This is where the advantages of neural networks come into play, as we can use transfer learning (Weiss et al., 2016) to train small amounts of data from multiple submesoscale active regions to fine-tune the model weight, and the cost is greatly reduced. Second, the spatial scale of the filter used in this study is 20km. The filtering scale and method will have a certain impact on the results. In the future, how to decompose the submesoscale and mesoscale signals in the data is also a problem worthy of in-depth study, which may be more important than the choice of deep learning algorithm (Srinivasan et al., 2023). Third, because the current training data is located in a small domain and the change in $f$ is small, $f$ is not added to the predictor. In the future, we will incorporate more training data with transfer learning and test it in a region with a larger meridional range.

The deep neural networks we use here are stable but simple in structure. This model can only calculate SVHF at a single moment and a single point, and lacks the consideration of the spatiotemporal correlation between variables. In the future, we will use Convolutional Neural Networks, Long Short-Term Memory, and other algorithms to further improve the parameterization. At present, most advanced algorithms are based on Python. It is very difficult to migrate them to the ocean model written by Fortran, which needs to be realized with the third-party libraries (C. Zhang et al., 2023), such as Fortran-Keras Bridge (FKB, Ott et al., 2020) . This may increase the computing cost, which needs to be considered.

**Acknowledgments**

The authors would like to thank Dr. Xingliang Jiang, Dr. Qingyue Wang, and Zhiwei You for their helpful discussions. This research was funded by the National Natural Science Foundation of China grant number 42176023, the National Key Research and Development Program of China, grant number 2021YFC3101601, and the National Natural Science Foundation of China, grant number 42176019.